\documentclass{emulateapj}
\usepackage{natbib}
\usepackage[caption=false]{subfig}
\usepackage{color}

\begin{document}
%\title{Predicting Shepherding Planet Properties from Debris Disk Observables}
%\title{Debris Disks and Planets: Constraining Planet Parameters from Disk Optical Depth Profiles}
%\title{Empirical Constraints on Planets Shepherding Debris Disks from Scattered Light Observables}
\title{Predictions for Shepherding Planets in Scattered Light Images of Debris Disks}
\author{Timothy J. Rodigas\altaffilmark{1,2}, Renu Malhotra\altaffilmark{3}, Philip M. Hinz\altaffilmark{1}}

\altaffiltext{1}{Steward Observatory, The University of Arizona, 933 N. Cherry Ave., Tucson, AZ 85721, USA; email: rodigas@as.arizona.edu}
\altaffiltext{2}{Carnegie Postdoctoral Fellow; Department of Terrestrial Magnetism, Carnegie Institute of Washington, 5241 Broad Branch Road, NW, Washington, DC 20015, USA}
\altaffiltext{3}{Lunar and Planetary Laboratory, University of Arizona, Kuiper Space Sciences Building, Tucson, AZ 85721, USA}

\newcommand{\about}{$\sim$~}
\newcommand{\mj}{M$_{J}$}
\newcommand{\degrees}{$^{\circ}$}
\newcommand{\arcseconds}{$^{\prime \prime}$}
\newcommand{\asec}{$\arcsec$}
\newcommand{\fasec}{$\farcs$}
\newcommand{\lprime}{$L^{\prime}$}
\newcommand{\ks}{$Ks$~}
\newcommand{\mjyasec}{mJy/arcsecond$^{2}$}
\newcommand{\microns}{$\mu$m}
\newcommand{\massratio}{$\mu/\mu_{J}$}

%\shorttitle{Predicting Planet Mass and Orbit from Disk Observables}
\shortauthors{Rodigas, Malhotra, $\&$ Hinz}

\begin{abstract}
Planets can affect debris disk structure by creating gaps, sharp edges, warps, and other potentially observable signatures. However, there is currently no simple way for observers to deduce a disk-shepherding planet's properties from the observed features of the disk. Here we present a single equation that relates a shepherding planet's maximum mass to the debris ring's observed width in scattered light, along with a procedure to estimate the planet's eccentricity and minimum semimajor axis. We accomplish this by performing dynamical N-body simulations of model systems containing a star, a single planet, and a disk of parent bodies and dust grains to determine the resulting debris disk properties over a wide range of input parameters. We find that the relationship between planet mass and debris disk width is linear, with increasing planet mass producing broader debris rings. We apply our methods to five imaged debris rings to constrain the putative planet masses and orbits in each system. Observers can use our empirically-derived equation as a guide for future direct imaging searches for planets in debris disk systems. In the fortuitous case of an imaged planet orbiting interior to an imaged disk, the planet's maximum mass can be estimated independent of atmospheric models.
\end{abstract}
\keywords{methods: N-body simulations --- circumstellar matter --- planetary systems} 

\section{Introduction}
Circumstellar dust has been detected around several hundred main sequence stars, and more than 30 such systems have now been spatially resolved at visible, near-, and far-infrared wavelengths (http://circumstellardisks.org). In these so-called debris disk systems, the dust is inferred to be generated by collisions in planetesimal belts that are dynamically shepherded by planets, analogous to the solar system's asteroid and Kuiper belts. In a few systems, notably $\beta$ Pictoris, HR 8799, and perhaps Fomalhaut, spatially resolved debris disk structure as well as one or more planets have been detected \citep{betapic,hr87994thplanet,kalas,curriefomalhaut,galicherfomalhaut,kalasnewfomalhaut}, allowing more detailed study of planet-planetesimal-disk interactions and of planetary system formation and evolution.  

Currently most spatially-resolved debris disks do not also have associated detections of planets, but many debris disks do show \textit{signs} of shepherding planets, such as clumps, gaps, and sharp edges. Indeed the presence of a planet in the Fomalhaut system has been predicted based on the resolved disk's properties \citep{quillenfomalhaut}, and \cite{chiang} (hereafter C09) have carried out detailed dynamical modeling to constrain the planet's mass to $< $ 3 M$_{J}$ by consideration of the debris disk's observed properties. A faint object scattering star light has recently been imaged in the system \citep{kalas,curriefomalhaut,galicherfomalhaut}, but its true nature and origin remain ambiguous \citep{kalasnewfomalhaut}.

The Fomalhaut dynamical predictions were very useful, but they were customized to a specific debris disk. What predictions can be made in the general case? Several studies have examined how planet mass affects debris disk structure \citep{wyatthr4796,kuchner,quillenfomalhaut,chiang,collisiondynamics,smack}, but these are not readily adapted for the inverse problem, namely estimating the planet mass and orbit from a given set of debris disk observations.

The goal of the present paper is to provide observers with a ``rough guide'' for estimating the mass and orbit of a putative shepherding planet from scattered light observations of a debris disk. The procedure we present is obtained as follows: we carry out a suite of N-body numerical integrations consisting of a single planet interacting with an exterior ring of dust-producing planetesimals; the simulations cover a range of system parameters, and we obtain a suite of simulated debris disks. We calculate the optical depth profile (a proxy for surface brightness) and measure the normalized full-width half-maximum (FWHM) of each simulated disk. We then examine the relationships between the FWHM and the simulation input parameters. Specifically, we provide observers with a simple procedure to estimate the maximum mass of a putative shepherding planet, its orbital semimajor axis, and its eccentricity from surface brightness profiles of scattered light images of debris disks. In Section 2, we describe our methods. In Section 3, we present our results and predictions for planets in five imaged debris rings. In Section 4, we discuss the implications of our results and outline the observer's procedure for estimating a planet's maximum mass and orbit from scattered light images of debris disks. In Section 5 we summarize our main findings.

\section{Methods}
We adopt the hypothesis that the scattered light of the debris disk arises from stellar radiation scattered by dust grains produced in an underlying ring of dust-producing planetesimals--``parent bodies"--shepherded by a nearby perturbing planet orbiting interior to its inner edge. As in C09, we first numerically integrate the (massless) parent bodies to produce an ensemble of stable orbits. The remaining stable parent bodies are then assumed to ``release" dust grains of a prescribed size distribution. The dust-producing parent bodies are large enough that stellar radiation pressure can be ignored for them, which results in disks that are intrinsically narrower (C09; \cite{collisiondynamics,smack,fomalhautalma}). However, radiation pressure is not negligible for the much smaller dust grains since it spreads them onto wide orbits (C09). Because radiation pressure causes a radial acceleration that is inversely proportional to astrocentric distance, it is simple to model it as a fraction, $\beta$, of the stellar gravitational acceleration, where $\beta$ is a function of particle size and density \citep{wyatt1950}. In our numerical simulations, we only include bound particles ($\beta < 0.5$). We ignore Poynting-Robertson (PR) light drag and account for collisions by integrating the dust grains for the number of orbits that correspond to their collisional lifetimes. See the Appendix for a discussion and justification of these choices.

We use a fast N-body orbit integrator, which is based on the second order mixed-variable method of \cite{wisdom1991}. This code, written in FORTRAN, provides an order of magnitude increase in speed compared to conventional integrators while minimizing numerical losses in the constants of integration (energy, angular momentum, $\beta$-modified Jacobi constant). To illustrate, a simulation of 10,000 particles interacting with a star and a perturbing planet for 1000 orbits of the planet, and using an integration step size of 5$\%$ of the planet's orbital period, takes only \about 1 minute of wall clock time on a 2010 computer\footnote{Macbook Pro, 4 GB memory, 2.66 GHz Intel Core 2 Duo}. The entire suite of 160 independent N-body simulations required to vary all the input parameters in this study takes only a few hours of wall clock time to complete.

\subsection{Simulation parameters}
The parameters of interest in this study are: the inner edge of the parent body disk, a$_{inner}$; the initial width of parent body disk, $w_{pb}$; the initial eccentricity of the parent body disk, e$_{disk,i}$; and the longitude of periastron of the parent body disk, $\varpi_{pb}$. The perturbing planet's parameters are its mass, $m_{p}$, its orbital semimajor axis, a$_{p}$, eccentricity, e$_{p}$, and longitude of periastron, $\varpi_{p}$. 

\begin{table*}[t]
\centering
\caption[Dynamics Simulation Parameters]{Simulation Parameters}
\label{tab:params}
\begin{tabular}{c c c}
\hline
Parameter & Value & Description  \\
\hline 
$\mu/\mu_{J}$ & (0.3, 1.0, 3.0, 10.0) & planet mass ratio / Jupiter's mass ratio \\
e$_{p}$ & (0.0, 0.05, 0.10, 0.15, 0.20) & planet eccentricity \\
a$_{p}$ & see Table \ref{tab:stability} & planet semimajor axis \\
$\varpi_{p}$ & 0 & planet's longitude of periastron \\
e$_{disk, i}$ & (0.0, 0.05, 0.10, 0.15, 0.20) & initial parent body disk eccentricity \\
a$_{inner}$ & 1 & inner edge of parent body disk \\
$\varpi_{pb}$ & 0 & initial longitude of periastron of parent body disk  \\
$w_{pb}$ & 0 & initial width of parent body disk \\
$\beta$ & (0.0, 0.00625, 0.0125, 0.025, & radiation pressure force / gravity  \\ 
& 0.05, 0.1, 0.2, 0.4) &  \\
a$_{peak}$ & output & optical depth profile peak semimajor axis \\
a$_{1/2}^{in}$ & output & optical depth profile inner half-peak semimajor axis \\
a$_{1/2}^{out}$ & output & optical depth profile outer half-peak semimajor axis \\
e$_{disk,f}$ & output & final parent body disk eccentricity \\
nFWHM & output & normalized optical depth profile FWHM \\
%Number of test particles & 5000 & \\
%total integration (parent bodies) & 1000 orbits & \\
%total integration (parent bodies) & 100 orbits & \\
%integration time step & 5$\%$ & \\
\hline 
\end{tabular}
\end{table*}

In spatially resolved scattered light images of debris disks, if the disks are not too inclined relative to our line of sight, we can typically determine the following disk parameters, or observables: the semimajor axis, a$_{peak}$, the eccentricity, e$_{disk}$, and longitude of periastron, $\varpi_{disk}$, of the deprojected disk. In this study, we will also utilize the normalized FWHM (nFWHM) of the (radiation dilution-corrected) surface brightness profile of the deprojected disk, defined here as 
\begin{equation}
\label{eqn:fwhm}
\hbox{nFWHM} = \frac{\hbox{a}_{1/2}^{out} - \hbox{a}_{1/2}^{in}}{\hbox{a}_{peak}},
\end{equation}
where a$_{1/2}^{out}$ and a$_{1/2}^{in}$ are the outer and inner semimajor axis locations where the deprojected disk surface brightness is half of the peak surface brightness\footnote{The right-hand side of Eq. \ref{eqn:fwhm} is often expressed as $\Delta R / R$.}. Other works have utilized the sharpness of the inner edge of the disk as the key observable indicating the presence of a disk-shepherding planet. However, the inner edge sharpness suffers from a degeneracy in planet mass/semimajor axis (ie, high-mass planets far away from the disk can produce the same sharpness as low-mass planets close to the disk; see Fig. 3 in C09). The width of the disk, on the other hand, is much less affected by this degeneracy, as will be demonstrated in Section \ref{sec:results}. See Table \ref{tab:params} for a list and description of the relevant parameters utilized in this work. 

\subsection{Initial conditions}
We adopt units whereby the stellar mass M$_{*}$ and the universal constant of gravitation, $G$, are unity, and the inner edge of the initial parent body disk, a$_{inner}$ is adopted as the unit of length. In these units, a particle orbiting at a$_{inner}$ has a period of 2$\pi$. We simulate systems with planet mass $\mu$ in the range (0.3, 1.0, 3.0, 10.0) $\mu_{J}$, where $\mu_{J} = 9.55 \times 10^{-4}$ is the mass ratio of Jupiter relative to our Sun. Although smaller planet masses are likely present in debris disks, we do not simulate these cases for two reasons. First, \cite{collisiondynamics} showed that over a wide range of optical depths and for $\mu < 0.3$ $\mu_{J}$, dust grain collisions can wash out observable effects on debris disks. Second, direct imaging instrumentation is currently not capable of detecting planets less massive than \about a few \mj.

The parent bodies' semimajor axes are all initialized at a$_{inner}$ (ie, initially infinitesimally-narrow disks with $w_{pb} = 0$). We also tested disks with non-zero initial widths but found that these cases introduced a degeneracy with planet mass/semimajor axis combinations. For example, consider an imaged debris disk with a FWHM = 15$\%$. This disk could have been broadened from an initial width of 10$\%$ due to interactions with a distant, high-mass planet; or it could have been broadened from an initial width of 14$\%$ by a very low-mass, very nearby planet. An observer has no way of knowing which planet produced the observed disk without knowing the initial parent body disk width, which cannot be measured. If we instead consider initially infinitesimally-narrow rings, then for a given observed disk width, we can at least constrain the \textit{maximum} mass of the putative planet, even if nature is unlikely to produce such initial parent body disks.

The star, planet, and parent bodies all orbit in the same plane so that their relative inclinations are zero. This assumption is made for simplicity and should not seriously change the results since most debris disks are thought to be flat (aspect ratio, defined as disk height divided by disk radius, typically on the order of a few percent). 

We consider five values of the initial eccentricity of the parent body disk, e$_{disk,i}$: (0, 0.05, 0.1, 0.15, 0.2). These values adequately span known debris disk eccentricities. The planet and the parent bodies all have the same initial longitude of periastron, which we initialize to zero. The parent bodies have random initial mean anomalies uniformly spaced between 0 and 2$\pi$.

We set the planet's eccentricity to be equal to the parent body disk's initial eccentricity. While other studies have used the forced eccentricity relationship between the planet and the dust (e.g., \cite{quillenfomalhaut} and C09), this relationship is derived from linear secular theory, which is only valid for small mass ratios ($\mu \lesssim 10^{-3}$) and low eccentricities \citep{mustill09}. Since the parent bodies we simulate are effectively indestructible over the simulation timescale (see the Appendix), assuming that they can acquire forced eccentricities from nearby perturbing planets leading up to the start of our simulations is reasonable. Moreover the relationship between the planet's and disk's eccentricity that we simulate is merely the initial relationship; the eccentricities of the parent bodies evolve over time. 

\subsubsection{Numerical determination of planet semimajor axis}
\label{sec:widths}
The initial semimajor axis of the planet is determined by means of a bootstrap procedure in which we numerically determine the inner edge of stable orbits exterior to the planet's orbit for the ranges of planet mass and eccentricity of interest. Although we could have made use of published formulas for the chaotic zone of a planet (e.g., \cite{wisdom1980}; C09; \cite{mustillchaotic}), the results in the literature do not adequately cover the range of planet mass and eccentricity of interest in our study. Therefore we determine the semimajor axis of the planet relative to the inner stable edge of the parent body disk as follows: we place a planet of a given mass at a$_{inner}$ and place parent bodies at discrete semimajor axis locations beyond a$_{inner}$, starting far from the planet. The parent bodies have the same eccentricity as the planet and have random mean anomalies between 0-2$\pi$. The system is then integrated for 1000 orbits of the planet. In this simulation and all others described below, we use an integration step size of 5$\%$ of the orbital period of a particle with semimajor axis = a$_{inner}$. Parent bodies that approach within the planet's Hill radius or cross within 0.1 a$_{inner}$ of the star or beyond 100 a$_{inner}$ are discarded. The width of the unstable zone is determined as the distance between the planet and the closest semimajor axis at which at least 90$\%$ of the parent bodies remain at the end of the integration. The final planet semimajor axis locations are reported in Table \ref{tab:stability}.

The large separations between the planet and disk ($\gtrsim$ 3.5 Hill radii) justify the assumption that the parent bodies in the disk are massless. At such large separations, the planet and disk are effectively ``decoupled" such that even if the disk is massive, the migration rate is slow over the timescale of the simulation (\cite{bromleymigration}; see the Appendix for more details).

We should not expect the widths of our chaotic zones to agree with those from previous studies (e.g., 2.0$\mu^{2/7}$ from C09, or 1.8$\mu^{1/5}$e$_{p}^{1/5}$ from \cite{mustillchaotic}) because we are testing larger planet masses and larger eccentricities. Additionally, the relationship between the planet's and disk's eccentricity in our study is different from previous works that typically use the forced eccentricity relationship. We also used a different stability criterion ($> 90\%$ of particles must survive the integration) to determine the chaotic zone widths, which may be more stringent than previous works (e.g., C09). 
\begin{table}[h]
\centering
\caption[Numerically determined \MakeLowercase{a$_{p}$/a$_{inner}$}]{Numerically determined \hbox{a$_{p}$/a$_{inner}$}}
\label{tab:stability}
\begin{tabular}{c|c c c c}
% a$_{p}$/a$_{inner}$
e$_{disk, i}$ & \massratio = 0.3 & \massratio = 1  & \massratio = 3 &  \massratio = 10  \\
\hline
0   &   0.86  &   0.80   &   0.71   &   0.59  \\
 0.05   &  0.86   &   0.80   &   0.70   &   0.55 \\
 0.10    & 0.86   &   0.80    &  0.69    &  0.52 \\
 0.15    &  0.86  &    0.79   &   0.69    &   0.49 \\
  0.20   & 0.84   &   0.79    &   0.67    &  0.47 \\
\hline 
\end{tabular}
\end{table}

From Table \ref{tab:stability}, the initial eccentricity of the disk widens the unstable zone only for the largest mass ratios. However, the initial eccentricities of debris disks cannot be measured. Therefore we cannot compute a master chaotic zone equation from the values in Table \ref{tab:stability}. Instead, because we are attempting to constrain the \textit{maximum} mass of a disk-shepherding planet in this study, we should only constrain the \textit{minimum} semimajor axis of a putative planet. Therefore we adopt as our unstable zone equation the power-law fit to the e$_{disk,i} = 0.20$ values, since these correspond to the widest chaotic zones and the planets that are the farthest from the disk's inner edge:
\begin{equation}
\label{eqn:chaotic}
{a}_{p} = {a}_{inner} / \left(1 + 10.23 \mu^{0.51}\right).
\end{equation}
Note that this equation is not used directly in our simulations, as the planet semimajor axes are explicitly determined in Table \ref{tab:stability}. Rather, this equation is to be used by observers when conducting the procedure outlined in Section \ref{sec:procedure}.

To review, we have two free parameters that will be varied in the simulations: $\mu$ and e$_{disk, i}$. a$_{p}$ is numerically determined, e$_{p}$ $=$ e$_{disk, i}$, and $\varpi_{p} = \varpi_{pb} = 0$. The simulations will produce two outputs: nFWHM, calculated using Eq. \ref{eqn:fwhm}, and e$_{disk,f}$, the final eccentricity of the simulated disks.

\subsection{Procedure}
With all parent body and planet input parameters determined, we integrate the system of the star, planet, and 5000 parent bodies for 1000 orbital periods of a particle at a$_{inner}$. After 1000 orbits, the parent bodies ``release" dust grains that have the same positions and velocities as their parents, as in C09. The dust grain orbits are then numerically integrated, accounting for the effects of radiation pressure by multiplying the stellar mass M$_{*}$ by $1-\beta$. 

As in C09, we simulate 8 different values of $\beta$; including $\beta = 0$, these are: (0.0, 0.00625, 0.0125, 0.025, 0.05, 0.1, 0.2, 0.4). We integrate each system with a given $\beta$ for 100 orbit periods of a particle at a$_{inner}$. The length of this integration is approximately the collisional lifetime of a dust particle in the typical debris disks we are simulating (see the Appendix, Eq. \ref{eqn:tcollision}). 

\subsubsection{Optical depth profiles}
To obtain the optical depth profiles of the simulated debris disks, we follow the procedure outlined in C09. We take each surviving dust particle's final Cartesian positions and velocities, construct a grid of concentric ellipses with eccentricity given by the final eccentricity of the parent body disk, e$_{disk, f}$, and ellipse center given by a$_{disk}$e$_{disk, f}$, and count the total number of surviving particles in a given ellipse. Here, e$_{disk, f}$ and a$_{disk}$ are the locations of the peak of the eccentricity and semimajor distributions of the surviving parent bodies, respectively.

To increase the signal-to-noise (S/N), we ``spread'' each dust particle out along its orbit (``Gaussian wire method") as in C09. In this method, a surviving particle is cloned and placed at discrete locations along its orbit; the orbital elements are determined from the position and velocity of the particle at the end of the integration and account for the effects of radiation pressure. The number of particle clones generated per orbit is chosen so that the total number of particles equals 10$^{6}$. For example, if 5000 particles survive, then each particle would be ``spread'' along its orbit at 200 locations equally spaced in true anomaly. For the $\beta > 0$ particles, each particle has a weight that is inversely proportional to its velocity at the end of the integration, effectively causing slower moving particles to create greater dust density. 

\begin{figure}[h]
\centering
\subfloat[]{\label{fig:betaprofiles}\includegraphics[scale=0.4]{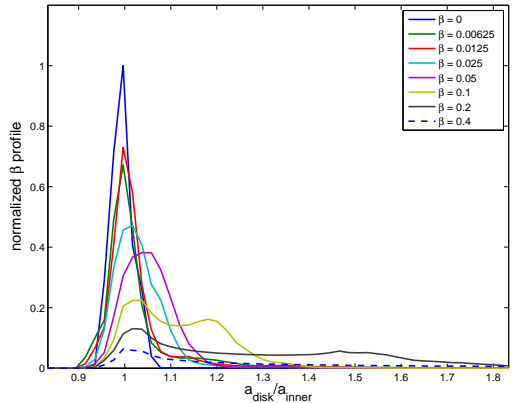}} \\
\subfloat[]{\label{fig:massprofiles}\includegraphics[scale=0.47]{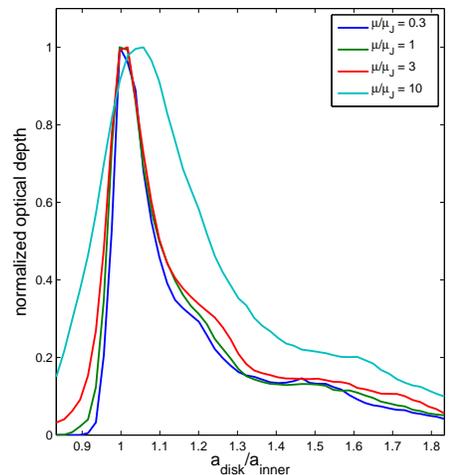}}
\caption[$\beta$ profiles for $\mu = 0.3 \mu_{J}$ and initial e$_{disk, i}$ = 0.10, and $\beta$-summed optical depth profiles for different $\mu$ values, all with e$_{disk, i}$ = 0.10]{\textit{Top}: $\beta$ profiles for $\mu = 0.3 \mu_{J}$ and initial e$_{disk, i}$ = 0.10. The profiles are very similar to what C09 observe for $\mu = 0.44 \mu_{J}$, e$_{p}$ = 0.12, and parent body disk width = 10$\%$. \textit{Bottom}: $\beta$-summed optical depth profiles for different $\mu$ values, all with e$_{disk, i}$ = 0.10. As the mass ratio increases, the profiles spread out, as is observed by C09 for their Fomalhaut simulations.}
\end{figure}

The number of ellipses used to construct the optical depth profile--which determines the ``resolution" of the final profile--is 50, with the first ellipse at 0.83 a$_{inner}$ and the last at 1.83 a$_{inner}$. This results in an optical depth resolution of 0.02 a$_{inner}$ per ellipse. This is lower than the resolution used in C09 (200 ellipses), but is well-matched to current observations. For example, if a$_{inner}$ = 50-100 AU (most debris disks reside at these separations), and 1 ellipse = 1 resolution element, then the resolution is 1-2 AU. This is comparable to the typical resolution achieved by HST for debris disks 50-100 pc from Earth.

For a given planet mass and disk eccentricity, we produce the final optical depth profile $\tau_{\perp}$ in the same manner as C09, by linearly combining the 8 different optical depth profiles for each $\beta$:
\begin{eqnarray}
\tau_{\perp} = \displaystyle\sum_{\beta \neq 0} N_{\beta} \frac{max(N_{0.00625})}{max(N_{\beta})} \left(\frac{\beta}{0.00625}\right)^{q-3} + \\ 
N_{0} \frac{max(N_{0.00625})}{max(N_{0})} (1 + \sqrt{2}), \nonumber
\end{eqnarray}
where $N_{\beta}$ refers to the optical depth profile for a specific $\beta$, $q$ is the differential power-law index assuming a collisional cascade in the disk and here assumed to be 3.5 \citep{dohnanyi}, and the $1 + \sqrt{2}$ term comes from the choice of binning (see C09, section 3.1.3 for a more in depth discussion of this constant term). 

\subsection{Control simulation}
Before running a full suite of N-body simulations, we verified that our simulations were producing results similar to those obtained by C09. While their input parameters (mass ratio, planet/disk eccentricity, initial parent body disk width) differ from ours, the general set up and methodology are very similar. Therefore we should expect to see similar results for similar inputs.

Fig. \ref{fig:betaprofiles} shows our $\beta$ profiles for the specific case of $\mu/\mu_{J}$ = 0.3 and initial disk eccentricity = 0.10. The profiles are very similar to Fig. 2 in C09. This gives us confidence that our simulations will yield accurate results, despite our differences relative to C09. 

We also verified that the perturbing planet was having the expected effect on the dust particles, namely spreading them out, resulting in wider optical depth profiles with increasing mass (as was seen by C09). Fig. \ref{fig:massprofiles} shows such an example simulation for an initial disk eccentricity of 0.10. The expected behavior is observed, again validating the methods and parameters chosen for the simulations. While the disk width increases only marginally for small mass ratios, the difference is readily evident when compared to the 10 \massratio ~model. This is satisfactory for the purposes of our study: discriminating between low-mass (\about 1 \massratio) and high-mass (\about 10 \massratio) planets in a given debris disk system.

\section{Results}
\label{sec:results}
\begin{figure}[h!]
\centering
\centering
\includegraphics[scale = 0.45]{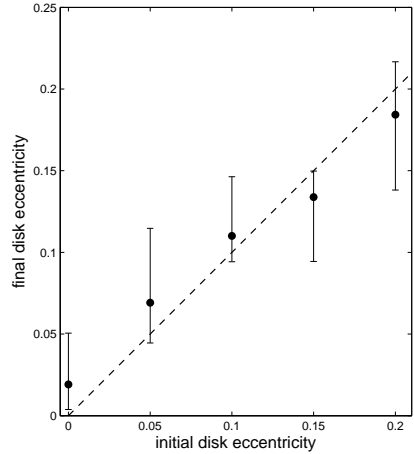}
\caption[Final disk eccentricity as a function of initial disk eccentricity for the various input planet masses]{Final parent body disk eccentricity as a function of initial parent body disk eccentricity for the various input planet masses. The points represent the averages of the final disk eccentricity values across mass, and the error bars represent the full range in values. The dashed line denotes perfect agreement between initial and final eccentricity. The introduction of a massive perturbing planet into the system can result in both an increase and a decrease in the disk's eccentricity, depending on the mass and initial eccentricity.} 
\label{fig:evse}
\end{figure}

\begin{figure}[h]
\centering
\includegraphics[scale = 0.45]{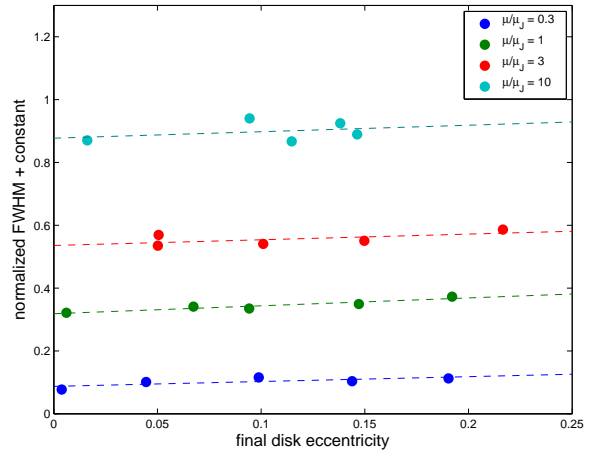}
\caption[Normalized disk FWHM as a function of final disk eccentricity for the various input planet masses]{Normalized disk FWHM as a function of final disk eccentricity for the various input planet masses. The different colors are each associated with a different mass ratio planet, and constant terms have been added to the FWHM values for graphical clarity. The dashed lines are fits to the data for each mass. There is no strong correlation between a debris disk's width and its eccentricity. This implies that disk eccentricity may not be a good indicator of a nearby perturbing planet.}
\label{fig:fwhmvse}
\end{figure}

After running the full suite of simulations, we measured $a_{1/2}^{out}$, $a_{1/2}^{in}$, and $a_{peak}$ for each $\beta$-summed optical depth profile (as in Fig. \ref{fig:massprofiles}) and computed the normalized FWHM of each modeled disk using Eq. \ref{eqn:fwhm}. We then examined the relationships between this output and the planet mass ratio and final disk eccentricity. We cannot use the initial disk eccentricity as an independent variable because it can increase or decrease from its original value, depending on the mass of the perturbing planet (see Fig. \ref{fig:evse}). 
Therefore we first examine the relationship between each disk's normalized FWHM and its final disk eccentricity (see Fig. \ref{fig:fwhmvse}). From Fig. \ref{fig:fwhmvse}, there is no strong correlation between these two variables. This means that disk eccentricity may \textit{not} be an indicator of a nearby massive planet (neglecting the possible dynamical interactions that could have excited the disk into an eccentric state in the first place). 

\begin{figure}[h]
\centering
\includegraphics[scale = 0.45]{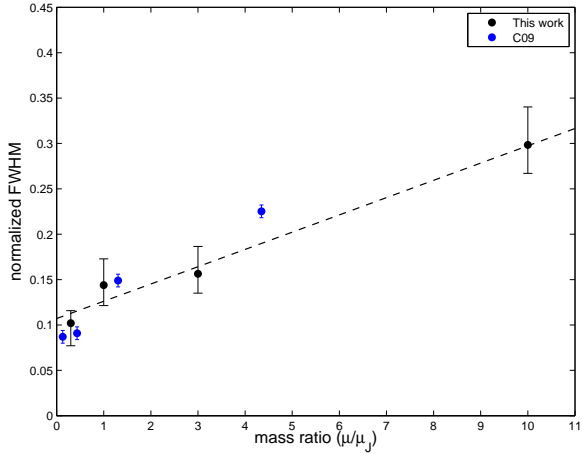}
\caption[Normalized disk FWHM as a function of planet mass ratio]{Normalized disk FWHM as a function of planet mass ratio. The black points represent the averages of nFWHM values at each mass for the various final disk eccentricities, and the error bars represent the full range in nFWHM values for each mass. The dashed line represents a linear fit to the points. The blue points and error bars represent equivalent values taken from Fig. 3 in C09. Increasing the mass ratio of the perturbing planet causes an increase in the range of possible disk FWHM values. A similar relationship was seen in C09 for Fomalhaut's disk.}
\label{fig:fwhmvsmass}
\end{figure}

Since the final disk eccentricity has little effect on the disk's width, we now examine the relationship between the normalized disk FWHM and planet mass ratio (see Fig. \ref{fig:fwhmvsmass}). Evidently increasing the mass ratio of the perturbing planet causes an increase in the range of possible disk FWHM values. 

Fitting a line to data yields an equation which can be used to estimate the maximum mass of a perturbing planet in a debris disk. To estimate the uncertainties in the slope and y-intercept, we fit all possible combinations of points and set the uncertainties as the differences between the minimum and maximum slope and y-intercept values. We then propagated these errors to obtain a total error for each predicted mass. Eq. \ref{eqn:fit} shows our linear fit to the data, along with the uncertainties in the slope and y-intercept: 
\begin{equation}
\label{eqn:fit}
\hbox{nFWHM} = (0.019 \pm 0.0064) \mu/\mu_{J} + (0.107 \pm 0.032).
\end{equation}
Interestingly, our fitted slope (0.019) is close to the slope from C09 (\about 0.03) obtained by fitting the normalized FWHM values from their Fig. 3 for the specific case of Fomalhaut. It is encouraging to see similar results despite the differences between our specific simulations (mass ratios, eccentricities, parent body disk widths, chaotic zone widths, and integration times). 

Inverting Eq. \ref{eqn:fit} and substituting terms, we now have an empirical estimate of the maximum mass of a disk-shepherding planet that depends solely on the disk's scattered light normalized FWHM: 
\begin{equation}
\label{eqn:mass}
m_{p}/\hbox{M}_{J} = \left(\frac{\hbox{nFWHM} - (0.107 \pm 0.032)}{0.019 \pm 0.0064}\right) \left(\frac{\hbox{M}_{*}}{\hbox{M}_{\odot}}\right).
\end{equation}

\subsection{Predictions for resolved debris rings}
We now use our empirically determined relationships to constrain the mass and orbit of shepherding planets in five bright ring-like debris disks. 
\subsubsection{Fomalhaut}
Fomalhaut is a very nearby A star with a debris disk detected in scattered light \citep{fomalhautkalas}. Recently a point source has been imaged orbiting interior to the ring \citep{kalas,fomalhautcurrie,galicherfomalhaut}. While originally posited as a planet shepherding the ring, it appears to be primarily scattering the star's light, its eccentricity is likely to be very large, and its orbit may intersect with the plane of the disk \citep{kalasnewfomalhaut}--all of which call into question its nature as a disk-shepherding planet. Nonetheless it is useful to estimate the maximum mass of a planet shepherding the disk, since such a planet may still exist in the system and future observations will seek to detect it. Since the disk's eccentricity is not a good predictor of planet mass, we need only the disk's deprojected normalized FWHM. This is \about 0.17 \citep{fomalhautkalas}. Inputting this value into Eq. \ref{eqn:mass} along with the star's mass of 2.3 M$_{\odot}$, we find a maximum planet mass of \about 7.6 $\pm 4.6$ \mj. The planet's eccentricity would be equal to the disk's (0.11), and its semimajor axis would be $\gtrsim $ 85 AU from Eq. \ref{eqn:chaotic}. Taking into account the inclination to the system, the minimum orbit-averaged projected separation would be \about 4\fasec 4. Recent imaging studies have already ruled out planets more massive than \about 1-3 \mj ~at these distances \citep{fomalhautjanson,currienewfomalhaut}. Therefore if a planet is currently shepherding the debris ring, it must be low-mass.                                                     

\subsubsection{HR 4796A}
HR 4796A has a bright debris disk that has been resolved at many wavelengths (e.g., \cite{hr4796schneider,thalmannhr4796,lagrange4796}). In scattered light, the disk appears as a narrow ring, with a large central gap between the disk and the star. The gap, the sharp inner and outer edges, and a small offset from the star are cited as evidence for a perturbing planet \citep{hr4796schneider,thalmannhr4796,wyatthr4796}. To date no planet has been detected, though \cite{lagrange4796} ruled out 3.5 \mj ~planets beyond 0\fasec 5 (36.5 AU projected separation).

According to \cite{hr4796schneider}, the disk's normalized FWHM is \about 0.18 from the deprojected radial surface brightness profile. With a stellar mass of 2.18 M$_{\odot}$, we estimate the maximum mass of a perturbing planet would be \about 8.4 $\pm 4.6$ M$_{J}$. The planet's eccentricity would be equal to the disk's (\about 0.05 from \cite{hr4796schneider} and \cite{thalmannhr4796}), and its semimajor axis would be $\gtrsim $ 45 AU. Its orbit-averaged minimum projected separation would be 0\fasec 14. Given the large shepherding planet mass this system can tolerate, along with the possibility that such a planet may have been missed by recent imaging campaigns \citep{thalmannhr4796,lagrange4796} due to its possible small projected separation, we advocate continued high-contrast imaging of this system in the coming years.

\subsubsection{HD 207129}
HD 207129 is a Sun-like star that has a large, faint debris disk, recently resolved in scattered light by \cite{kristhd207129} with HST. The disk has a normalized FWHM of \about 0.18. Taking a stellar mass of 1.1 M$_{\odot}$ \citep{kristhd207129}, we obtain using Eq. \ref{eqn:fit} a maximum planet mass 4.2 $\pm 2.3$ M$_{J}$. Since the eccentricity of the disk is only constrained to be $< 0.08$, the same constraint applies to the planet's eccentricity. Its semimajor axis would be $\gtrsim $ 92 AU, corresponding to a minimum orbit-averaged projected separation on the sky of \about 2\fasec 8. Despite this large separation, the star's old age (\about 1 Gyr from \cite{kristhd207129}) and small maximum planet mass render this system unfavorable for direct imaging planet searches.

\subsubsection{HD 202628}
HD 202628 is a Sun-like star with a very wide disk recently resolved by HST \citep{kristhd202628}. The disk has a normalized FWHM of \about 0.4 and an eccentricity of \about 0.18. Assuming a solar mass for the star, the maximum mass of a single perturbing planet would be \about 15.4 $\pm 5.5$ M$_{J}$. Its eccentricity would be 0.18, and its minimum semimajor axis would be \about 71 AU, corresponding to a minimum orbit-averaged projected separation of \about 1\fasec 2. Despite the star's likely old age (2.3 Gyr; \cite{kristhd202628}), its broad disk can tolerate a very massive planet that would likely still be detectable by current direct imaging technology. Therefore we advocate high-contrast imaging of this system.

\begin{table}[h]
\centering
\caption[Dynamically Predicted Masses and Orbits for Known Debris Disks]{Predicted Masses and Orbits}
\label{tab:masses}
\begin{tabular}{c c c c c}
\hline
Star & m$_{p}$/M$_{J}$ & a$_{p}$/AU & proj. sep. (\asec)$^*$ & e$_{p}$  \\
\hline
Fomalhaut & $<7.6 \pm 4.6$ & $>85$ & $>4.4$ & $0.11$ \\
HR 4796A & $<8.4 \pm 4.6$ & $>45$ & $>0.14$ & $0.05$ \\
HD 207129 & $<4.2 \pm 2.3$ & $>92$ & $>2.8$ & $< 0.08$ \\
HD 202628 & $<15.4 \pm 5.5$ & $>71$ & $>1.2$ & $0.18$ \\
HD 181327 & $<15.2 \pm 5.6$ & $>35$ & $>0.55$ & $0$ \\
\hline 
\end{tabular}
\raggedright
$^*$Orbit-averaged
\end{table}

\subsubsection{HD 181327}
HD 181327 is a Sun-like star with a large, bright debris disk resolved by HST \citep{hd181327schneider}. The most current analysis of the resolved images reveals that the disk has a normalized FWHM of 0.32 \citep{hd181327ice} and an eccentricity consistent with zero (C. Stark, private communication). For a stellar mass of 1.36 M$_{\odot}$ \citep{hd181327ice}, the shepherding planet's maximum mass would be \about 15.2 $\pm 5.6$ M$_{J}$. Its eccentricity would be \about 0 and its minimum semimajor axis would be \about 35 AU, corresponding to a minimum orbit-averaged projected separation of \about 0\fasec 55. \cite{wahhaj} imaged this star as part of the Gemini NICI Planet-Finding Campaign and ruled out planets more massive than \about 6 \mj ~beyond 0\fasec 36, effectively ruling out a high-mass perturbing planet. Therefore if this system contains a solitary shepherding planet, it must be low-mass.

See Table \ref{tab:masses} for a summary of the predicted masses and orbits for planets in each system. 

\subsection{Parent body disk widths}
\begin{figure}[h!]
\centering
\includegraphics[scale = 0.45]{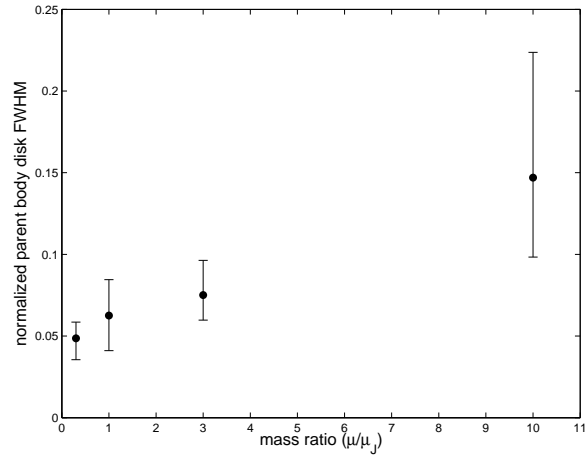}
\caption[Normalized parent body disk FWHM as a function of planet mass ratio]{Normalized parent body disk FWHM as a function of planet mass ratio. The black points represent the averages of the values at each mass for the various disk eccentricities, and the error bars represent the full range in nFWHM values for each mass. While a linear trend is evident, there is much more scatter than in the plot relating the dust disk FWHM to the planet mass ratio (Fig. \ref{fig:fwhmvsmass}), making the predictive power less precise. However, we can infer that parent body disk widths $< 10\%$ cannot contain a planet with mass ratio $\gtrsim$ 10 \massratio.}
\label{fig:pbfwhm}
\end{figure}

The predictions for planet mass and orbit in this study make use of the dynamical effects on dust grains. Are there any dynamical signatures on the parent bodies that produce the dust? Fig. \ref{fig:pbfwhm} shows the final normalized parent body disk width as a function of planet mass ratio. Clearly there is more scatter such that degeneracies in mass exist for widths $\lesssim 10\%$. However, we can still determine that parent body disks with widths $\lesssim 10\%$ cannot contain planets with mass ratios $\gtrsim$ 10 \massratio. While not as strong a constraint as can be made with scattered light images of a disk's dust, this is still useful in ruling out very massive planets in narrow parent body disks.

\section{Discussion: Observer's Procedure}
\label{sec:procedure}
Observers can use the results of this study in three ways. \textbf{(i)} For debris disks that have been resolved in scattered light but in which no planets have been detected, they can estimate the maximum mass, minimum semimajor axis, and eccentricity of a putative solitary shepherding planet. These can be calculated using the following procedure: 
\begin{enumerate}
\item Calculate the eccentricity of the debris disk, e$_{disk, f}$, from the deprojected scattered light image. The planet's eccentricity can be assigned this value.
\item Construct the azimuthally-averaged radial profile of the disk's deprojected scattered light, multiply the profile by the distance from the star squared to account for the geometric dilution of star light, and normalize the profile by the peak value.
\item Calculate the semimajor axis of the peak in the deprojected azimuthally-averaged radial profile, and the two locations equal to half the peak (a$_{peak}$, a$_{1/2}^{in}$, a$_{1/2}^{out}$). Use Eq. \ref{eqn:fwhm} to calculate the disk's normalized FWHM.
\item Insert the disk's normalized FWHM into Eq. \ref{eqn:mass} to solve for the maximum mass of the perturbing planet.
\item Assume a$_{1/2}^{in}$ = a$_{inner}$ and insert this value, along with the calculated maximum mass of the planet, into Eq. \ref{eqn:chaotic} to solve for the planet's minimum semimajor axis.
\end{enumerate}

This procedure (and for example, Fig. \ref{fig:massprofiles}) requires that debris rings be detected at high S/N in scattered light. This should be feasible in the coming years since HST/STIS is capable of detecting bright ring-like disks at very high S/N (G. Schneider, private communication; C. Stark, private communication). Additionally, ground-based telescopes with adaptive optics (AO) systems are now detecting disks in scattered light at high S/N \citep{esposito32297,thalmannhr4796,lagrange4796,hip79977,currie32297,boc32297,mehd15115,moth}. The visible light camera, VisAO \citep{visao}, on the Magellan AO system (MagAO; \cite{magao}) may be capable of producing the highest resolution images yet on known bright disks (Rodigas et al. 2014, in prep.), allowing for more precise measurements of scattered light disk widths. 

\textbf{(ii)} While no such system exists yet, for resolved debris disks that also contain a directly imaged planet, observers can use the above procedure to place an atmospheric model-independent limit on the planet's mass, semimajor axis, and eccentricity. This is especially useful because the masses of directly imaged planets depend heavily on atmospheric models (e.g., \cite{baraffe,burrows}), which themselves also depend on the (usually uncertain) age of the host star and on the initial conditions of the planet's formation (ie, ``hot-start" vs. ``cold start"; \cite{spiegel12}). 

\textbf{(iii)} For debris disks with information only available for the parent bodies, rather than the dust, observers can use Fig. \ref{fig:pbfwhm} to constrain the maximum disk-shepherding planet mass. For example, models of an unresolved disk detected by its infrared excess can sometimes estimate the location and width of the parent body disk. This information can be compared with our results to constrain a putative shepherding planet's mass. Additionally the Atacama Large Millimeter Array (ALMA) is now making it possible to directly image the tracers of parent bodies in debris disks (e.g., \cite{fomalhautalma}), which can provide model-independent measurements of parent body disk widths for comparison with our results. 

\section{Summary}
Using N-body simulations consisting of a star, a planet, and a disk of parent bodies and dust grains, we have shown that the width of a debris disk in scattered light is proportional to shepherding planet mass. This relationship can be used to estimate the maximum planet mass in a debris disk of a given width, as well as the planet's eccentricity and minimum semimajor axis.

Using the procedure outlined above, we have estimated the masses and orbits of putative shepherding planets in five bright debris rings. For Fomalhaut, deep imaging has already yielded mass limits below the maximum masses we report here. For HR 4796A, our results indicate a massive planet might reside close to the star, evading previous imaging detection efforts. For HD 207129, despite the favorable separation between the star and putative planet, the low maximum planet mass and old age make this system unfavorable for direct imaging. HD 202628 can contain a very high-mass planet at a favorable projected separation, making it attractive for direct imaging planet searches. While HD 181327 can also potentially host a high-mass perturber, \cite{wahhaj} have already ruled out such planets, implying that if the disk is being shepherded by a planet, the planet must be low-mass.

In general, observers searching for planets should prioritize systems that contain wide debris disks, as these can tolerate more massive interior perturbers. Once a planet is directly detected orbiting interior to its disk, its mass can be estimated independent of atmospheric models, providing a check on this fundamental physical property. 

\acknowledgments
We thank the anonymous referee for very helpful comments which greatly improved this paper. We thank Eugene Chiang for helpful discussions, and for sharing data from \cite{chiang}. We thank Andy Skemer, John Debes, and Chris Stark for helpful discussions. T.J.R. was supported by the NASA Earth and Space Science Fellowship (NESSF). 

\appendix
\section{Assumptions, forces, and timescales}
\subsection{Parent bodies}
We make several simplifying assumptions regarding the parent bodies to facilitate the many simulations we carry out. First, we assume that all particles have zero mass (as in C09). Doing so allows us to treat each parent body as a test particle interacting purely gravitationally with a massive planet and central star (ie, the restricted 3-body problem), which is much less computationally strenuous. 

The assumption that the parent bodies have zero mass is of course not realistic. The masses of debris disks are not well constrained, but a few model-dependent measurements of the mass of Fomalhaut's debris disk estimate a total mass of \about 1-110 M$_{\oplus}$ (C09; \cite{fomalhautalma,fomalhautherschel}), where M$_{\oplus}$ is the mass of Earth. If we conservatively assume that our simulated parent body disks have masses at the upper end of this range, what happens to the planet? \cite{bromleymigration} showed that for a single planet orbiting interior to a non-zero mass parent body disk, the planet will migrate away from the disk as long as the separation is $\gtrsim 3.5$ Hill radii. At such large separations, the migration rates are slow \citep{bromleymigration}. Using their analytic equations, after 10,000 orbits a 1 M$_{J}$ planet will have migrated away from a 100 M$_{\oplus}$ parent body disk of 10$\%$ width by only \about a few percent of its original semimajor axis. More massive planets will migrate even less \citep{bromleymigration}. Furthermore these results assume a constant migration rate over time, which is unlikely. Therefore neglecting migration and assuming zero mass for the parent bodies is reasonable for our purposes.

Second, we assume (as in C09) that over the course of the numerical integrations, the parent bodies are not destroyed by collisions, and their orbits are unaffected by collisions with smaller bodies. The collisional lifetime of a particle in a debris disk, or time before a particle is catastrophically destroyed via collision with a similarly-sized particle, can be written as 
\begin{equation}
\label{eqn:colltime}
t_{collision} = {t_{per}\over\pi \tau_{\perp} f},
\end{equation}
where $t_{per}$ is the orbital period of the particle in the disk, $\tau_{\perp}$ is the disk's vertical optical depth, and $f$ is a factor that depends on the particle size \citep{wyatt99}. For a typical dust grain, $f$ \about 4 and its collisional lifetime in a Fomalhaut-like debris disk is only \about 10$^{5}$ years (see the following subsection). From \cite{wyatt99}, the collisional lifetime of a parent body at the top of the collisional cascade can be written as
\begin{equation}
\label{eqn:colltimepb}
t_{collision, pb} \approx t_{collision} \left({D_{pb}\over D_{dust}}\right)^{1/2},
\end{equation}
where $D_{pb}$ and $D_{dust}$ are the diameters of the parent bodies and dust grains, respectively, and a \cite{dohnanyi} size distribution has been assumed. For a Fomalhaut-like debris disk, a typical dust grain size is \about 10 \microns ~(C09). Using the above dust grain collisional lifetime, Eq. \ref{eqn:colltimepb} shows that the collisional lifetime of parent bodies as small as just 10-100 m is \about 100-300 Myr, or \about the age of Fomalhaut. Such small parent body sizes are rather conservative based on models of extrasolar debris disks (C09, \cite{wyatt99}), as well as estimates of primordial parent body sizes in the solar system's asteroid and Kuiper belts \citep{asteroidsizes,kuiperbeltsizes}, which typically prefer sizes of \about 1 km or larger. Therefore it is reasonable to assume that the parent bodies in our study are effectively ``indestructible" on a timescale comparable to the collisional lifetime of the dust grains, but still produce dust via collisions with smaller particles. 

\subsection{Dust grains}
Because our study concerns debris disks (where little or no gas is present), the forces acting on dust grains are primarily: gravity from the host star, gravity from the perturbing planet, radiation pressure from the host star, and PR drag. In addition, the lifetimes of dust grains are limited by grain-grain collisions, for these lead to catastrophic fragmentation. For a geometrically absorbing dust grain of bulk density $\rho$ and radius $s$, the ratio, $\beta$, of the radial force of stellar radiation pressure to the force of stellar gravity is,  
\begin{equation}
\label{eqn:beta}
\beta= {3L_*\over 16\pi GM_*c\rho s}, %= 5.7\times10^{-5}(L_*/L_\odot)(M_*/M_\odot)^{-1}(\rho s)^{-1},
\end{equation}
where $L_*$ and $M_*$ are the stellar luminosity and mass, respectively, $G$ is the universal constant of gravitation, and $c$ is the speed of light. In response to radiation pressure, dust grains generated by parent bodies moving on circular orbits acquire larger, more eccentric orbits, with semimajor axis and eccentricity given by
\begin{equation}
a= a_{pb}(1-\beta)/(1-2\beta),\qquad e= \beta/(1-\beta).
\label{eqn:grainorbit}
\end{equation}
Only the particles with $\beta < 0.5$ remain bound to the star; in other words, bound particles are larger than the ``blow out'' size (C09), 
\begin{equation}
s_{b} = {3L_*\over 8\pi GM_*c\rho s}= 1.16 ({\rho\over1~\hbox{g~cm}^{-3}})^{-1} ({L_*\over L_\odot}) ({M_*\over M_\odot})^{-1}\mu{\rm m}.
\end{equation}
Smaller particles, $s<s_b$, acquire hyperbolic orbits upon release from the parent bodies, and their residence time in the debris disk is 
\begin{equation}
t_{unbound} \approx {w\over \Omega(a_{pb})}=16({w\over 0.1})({a_{pb}\over\hbox{100 AU}})^{3/2}({M_*\over M_\odot})^{-1/2}\hbox{yr},
\label{eqn:tb}
\end{equation}
where $\Omega(a) = \sqrt{GM_*/a^3}$ is the local Keplerian frequency and $w$ is the normalized width of the disk. (For parent bodies on moderately eccentric orbits, the orbit of the dust grain depends on the longitude at which it is released, and so does the minimum blow out size, but the latter is still close to $s_b$ given above.)
From \cite{wyatt1950}, for the bound dust grains, PR drag causes orbits to circularize on a timescale
\begin{eqnarray}
t_{PR} = & \left|{\dot e\over e}\right|^{-1} = {32\pi\rho s c^2 a^2\sqrt{1-e^2}\over 15 L_*} \nonumber \\
\approx & 7\times10^6{(1-\beta)\over \beta(1-2\beta)^{3/2}}({M_*\over M_\odot})^{-1}({a_{pb}\over100~\hbox{AU}})^2~\hbox{yr}; 
\label{eqn:tPR}
\end{eqnarray}
 the orbit decay timescale is $|\dot a/ a|^{-1} \approx 2(1-2\beta)t_{PR}$ for $\beta\lesssim 0.5$.

Eq. \ref{eqn:colltime}, the characteristic lifetime for destruction of grains by grain-grain collisions in the bound population, can be rewritten as,
\begin{eqnarray}
t_{collision} & \sim& \left({\tau_R \Omega(a)\over w}\right)^{-1} \nonumber\\
&\sim& 5\times10^4 ({10^{-3}\over\tau_R}) ({w\over0.1})^{1/2} ({a_{pb}\over100~\hbox{AU}})^{3/2}({M_*\over M_\odot})^{-1/2}~\hbox{yr},
\label{eqn:tcollision}
\end{eqnarray}
where $\tau_R$ is the radial optical depth, and $\tau_R/\sqrt{w}$ accounts for the effective optical depth for the path length $a_{pb}\sqrt{w}$ traversed by a dust grain on an elliptical orbit of eccentricity near unity. In Eqs.~\ref{eqn:tPR} and \ref{eqn:tcollision} we have also used the relation between the dust grain's orbital parameters and those of the parent bodies (Eq. \ref{eqn:grainorbit}).

Comparing these three timescales for dust grains, we see that for typical debris disks imaged in scattered light (ie, those with large optical depths), $t_{PR}\gg t_{collision} \gg t_{unbound}$ (as in C09, \cite{wyattprdrag}). This means that particles below the blowout size do not contribute to the brightness of a debris disk. Therefore we only simulate bound particles. Additionally, we can see that on timescales on the order of the dust grain lifetimes, $\sim t_{collision}$, the bound dust grains' orbits do not change significantly from their initial orbits upon ``release''. This justifies our neglect of PR drag in the simulations. 

\section{Limitations of this study}
Here we describe several limitations of our study that we have not addressed elsewhere. We sought dynamical stability of all parent bodies for only 10$^{3}$ orbits so that the wall clock time for a given simulation would be short. Integration times of $> 10^{6}$ orbits would have been more realistic since imaged debris disks can be as old as \about 1 Gyr. Nonetheless our shorter integrations provide a good approximation because test particle dynamical lifetimes near planets have approximately logarithmic distributions \citep{holman1993,minton2010} so that longer integration times will not significantly change the stable population.

We assumed that integrating the dust particles for 100 orbits was satisfactory for all debris disks. The number of dust grain orbits is directly related to the optical depth of the disk, and in our case, we simulated Fomalhaut-like debris disks. Observed debris disks have different optical depths, which means the amount of time before a dust grain is destroyed changes for each system. Our simulations do not account for this. However, we did test longer dust grain integrations (1000 orbits), corresponding to lower optical depths, and found this made little difference on the final results.

The assumption that all debris disks are collision-dominated is not realistic, since some disks are assuredly dominated by PR drag. However, to be PR drag-dominated, the optical depth in a disk must be very low \citep{wyatt99}. Such disks are likely to be very faint in scattered light and are consequently much harder to image. Therefore our neglect of such disks in our simulations is justified.

We assumed a differential power-law index for collisional cascades of 3.5, based on \cite{dohnanyi}. This is a common assumption but will certainly not be valid for all debris disks. \cite{andras} found evidence for the index being closer to 3.65, but this difference is small. Additionally, \cite{fomalhautq} recently found an index of \about 3.5 for Fomalhaut's disk, lending support to our choice.

%We have assumed that a scattered light debris disk surface brightness (SB) profile is equivalent to our modeled optical depth profiles, which is not in fact true. To obtain a vertical optical depth profile, one must account for the geometric dillution of star light by multiplying the SB profile by the distance from the star squared \citep{alycia141569}. Since most debris disks are narrow, we do not anticipate a large error in this assumption.

%We assumed that the initial parent body disk width was zero to remove degeneracies in this parameter. It is unlikely that parent bodies in debris disks are initially confined to infintesimally-thin rings. However, this assumption does not preclude us from estimating the maximum mass of a disk-shepherding planet. From our results, an observer can rule out planets above a certain mass that would have dynamically broadened a given disk more than is observed.

Finally, we stress that our models cannot explain all sharp, eccentric debris rings. Rather, the goal of this study was to answer the following question: \textit{if the observed disk features are being created by a single planet orbiting interior to the disk}, how massive is that planet and what is its orbit like? It is certainly possible that other physical processes (e.g., dust grain collisions, radiation forces, dynamical interactions with a distant star, the interstellar medium, multiple low-mass planets (e.g., \cite{raymonddisks,raymonddisks2}), or even gas \citep{kuchnergas}) could produce disk features that we assume are created by a solitary planet. 

%\clearpage

\bibliographystyle{apj}
\bibliography{ms}

%\bibliography{ms}

\end{document}